\documentclass[12pt]{iopart}

\usepackage{graphicx}
\usepackage{color}
\usepackage[toc,page]{appendix}
\usepackage{amsmath}
\usepackage{hyperref}
\newcommand\ringring[1]{%
  {% make an Ord atom
   \mathop{\kern0pt #1}\limits^{% set a box over the variable
     \vbox to-1.85ex{
       \kern-2ex % lower the ring accents
       \hbox to 0pt{\hss\normalfont\kern.1em \r{}\kern-.45em \r{}\hss}%
       \vss % fill
     }% end of \vbox
   }% end of the superscript
  }% end of \mathop
}

\begin{document}

%\title[Space-time approach to locomotion on curved substrate]{Space-time approach to locomotion on curved substrate}

\title{A robophysical model of spacetime dynamics}

% Space-time approach to locomotion in deformable environments
% Robophysical modeling of spacetime dynamics
% Programming active matter on curved terrain
% Programming active matter on elastic substrate

\author{Shengkai Li$^1$, Hussain N. Gynai$^2$, Steven Tarr$^2$, Emily Alicea-Mu\~noz$^2$, Pablo Laguna$^3$, Gongjie Li$^2$, Daniel I. Goldman$^{2*}$}
\address{$^1$ Department of Physics, Princeton University, Princeton, 08544, New Jersey, USA}
\address{$^2$ School of Physics, Georgia Institute of Technology, Atlanta, 30332, Georgia, USA}
\address{$^3$ Center for Gravitational Physics, Department of Physics, University of Texas at Austin, Austin, 78712, Texas, USA}

\ead{daniel.goldman@physics.gatech.edu}
\vspace{10pt}
\begin{indented}
\item[]Jan 2023
\end{indented}
%Dissipative forces cause the initially GR-like dynamics to be transient and consequently restrict experimental study to only the beginnings of trajectories; dominance of Earth's gravity forbids the difference between spatial and temporal spacetime curvatures
% Active matter on elastic substrate shares the same notion as GR. Inspired by GR, we embed the motion into a spacetime to obtain conserved quantities to help us understand the dynamics and further design the dynamics.
\begin{abstract}
Systems consisting of spheres rolling on elastic membranes have been used to introduce a core conceptual idea of General Relativity (GR): how curvature guides the movement of matter. However, such schemes cannot accurately represent relativistic dynamics in the laboratory because of the dominance of dissipation and external gravitational fields. Here we demonstrate that an ``active" object (a wheeled robot), which moves in a straight line on level ground and can alter its speed depending on the curvature of the deformable terrain it moves on, can exactly capture dynamics in curved relativistic spacetimes. Via the systematic study of the robot's dynamics in the radial and orbital directions, we develop a mapping of the emergent trajectories of a wheeled vehicle on a spandex membrane to the motion in a curved spacetime. Our mapping demonstrates how the driven robot's dynamics  mix space and time in a metric, and  shows how active particles do not necessarily follow geodesics in the real space but instead follow geodesics in a fiducial spacetime. The mapping further reveals how parameters such as the membrane elasticity and instantaneous speed allow the programming of  a desired spacetime, such as the Schwarzschild metric near a non-rotating blackhole. Our mapping and framework facilitate creation of a robophysical analog to a general relativistic system in the laboratory at low cost that can provide insights into active matter in deformable environments and robot exploration in complex landscapes.
\end{abstract}

%\maketitle

\section{Introduction}

% REMOVE THE PEDAGOGICAL. ACTIVE ALLOWS MAPPING TO GR.

Systems consisting of spheres rolling on curved surfaces \cite{middleton2016elliptical,white2013trajectories} are a well-known non-hydrodynamic analog to gravity. In such readily accessible systems, researchers have made intriguing connections to gravity such as Kepler-like laws, precession, and the stability of orbits. However, their studies have also found that these systems do not exactly mimic astrophysical gravity. For instance, the scaling between the period and radius is $T \propto r^{2/3}$ \cite{white2002shape} instead of $T \propto r^{3/2}$ in Kepler's third law. Additionally, the sphere on the elastic membrane is passive; as a result, not only do trajectories decay, but also the tunable parameters are limited to only the boundary conditions and the mass of the sphere.

% For instance, the scaling between the period and radius is either $T \propto r^{2/3}$ for light central mass or $T \propto r$ for heavy mass \cite{white2002shape}, instead of the $T \propto r^{3/2}$ in Kepler's law.

\begin{figure}[ht!]
  \centering
  \includegraphics[width=0.7\textwidth]{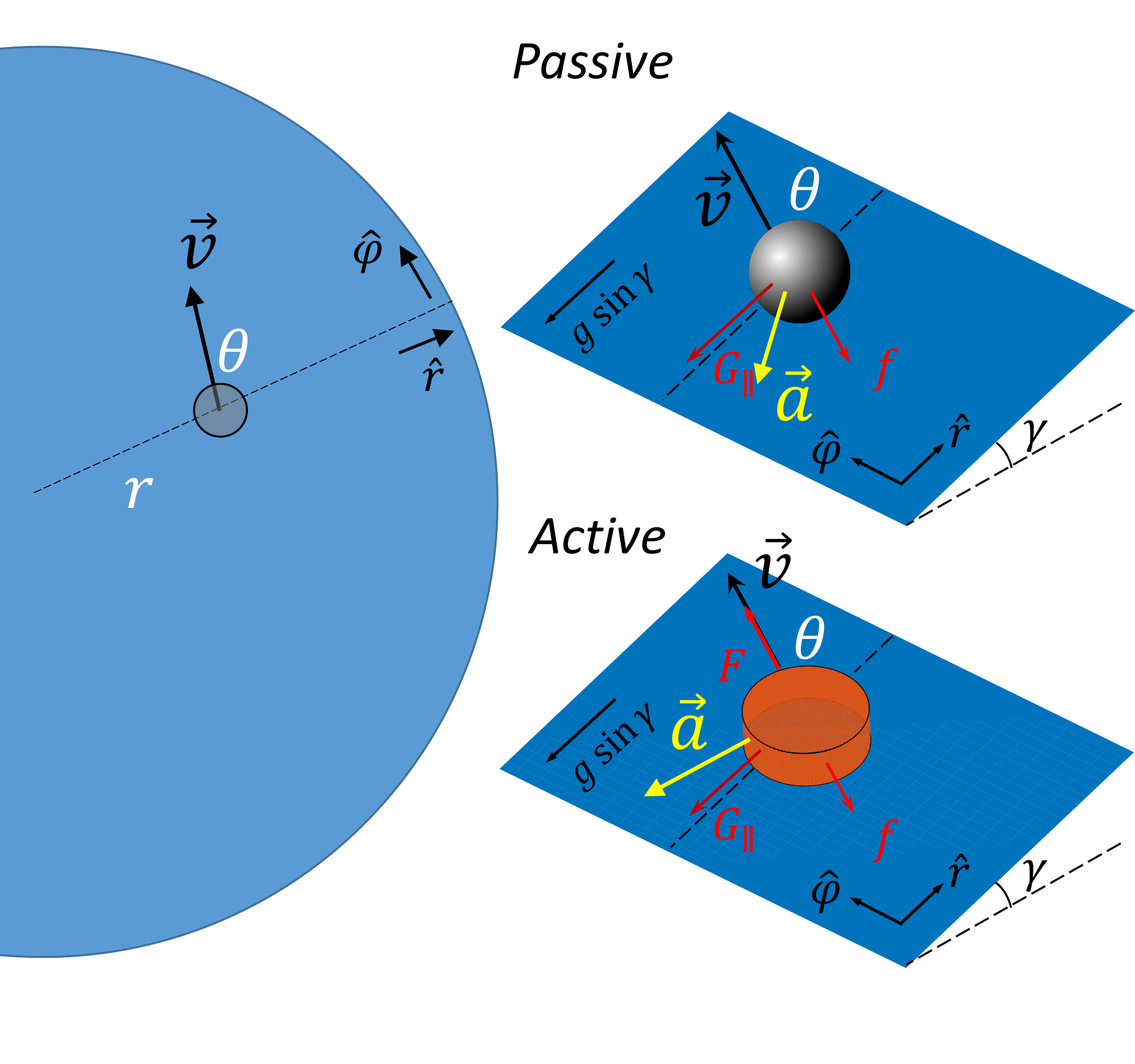}
  \caption{\textbf{A passive object (a marble) versus an active object (a robot) on a deformable membrane.} While a passive marble is only subjected to friction and Earth's gravity that leads to energy dissipation as $\vec{F}\cdot\vec{v}\propto\vec{a}\cdot\vec{v}<0$, an active object with an additional drive force can maintain steady-state motion with prescribed speed as $\vec{a}\cdot\vec{v}=0$.}
  %\vspace{-5mm}
  \label{fig:activePassive}
\end{figure}

We hypothesized that making the object ``active" -- an internally driven robot  -- would allow mechanical systems to better model GR in part because of the ability to study steady states. We further reasoned that the programmability and sensory capabilities of increasingly low-cost and powerful ``robophysical'' models \cite{aguilar2016review,aydin2019physics} could  allow the  tuning of parameters that lead to inexact mimics of GR in passive systems. Indeed, our recent work \cite{fieldMediated} built a framework to understand how the field-mediated dynamics of active agents on flexible membranes demonstrate in the words of Wheeler's famous aphorism: ``Matter tells spacetime how to curve and spacetime tells matter how to move"~\cite{wheeler1991geons}. In particular, we showed  that the spacetime followed by an active object can be tuned by varying system parameters such as the membrane elasticity and the speed of the object. 

Here we amplify on and extend the scheme introduced in~\cite{fieldMediated} and demonstrate how the activity can lead to an exact mapping to GR. We first show how an active object with a prescribed speed on an elastic membrane produces longer and more controllable trajectories compared with a passive marble. We then deduce the spacetime it follows, and subsequently show one can program the spacetime with a Schwarzschild orbit as an example. We posit that a future robotic vehicle controlled in the way we describe could mechanically mimic blackhole dynamics in the laboratory at a low cost.\\

\section{An active object with fixed speed on an elastic membrane}\label{sec2}

We first consider an active object prescribed with a constant speed on a circular elastic membrane. Later, we will discuss the general case of time-varying speed. To prevent the object from simply following a near-straight-line spatial geodesic with a spatial curvature
\begin{eqnarray}
ds^2=\Psi^2 dr^2+r^2 d\varphi^2 \label{eq:spaGeod}
\end{eqnarray}
where $\Psi^2=1+z'^2$ and prime denotes the derivative with respect to $r$, the object must  turn according to the instantaneous local slope, $-\nabla z$, the negative gradient of the terrain (membrane) height $z$. We enable this behavior by using a vehicle with a differential drive, a mechanism that permits a difference between the speeds of the vehicle's two wheels to respond to the terrain slope while maintaining a center-of-mass speed prescribed by the motor.  The vehicle drives straight on a leveled flat ground. When driving on a tilted flat surface with a constant gradient everywhere, the vehicle turns to align with this constant slope. 
In a more general case where the vehicle drives on a solid terrain with spatially varying gradient, it responds instantaneously by aligning with the local gradient. Finally, in our setup, the vehicle responds to the local gradient while the terrain (membrane) updates its shape with the evolving position of the vehicle,  the vehicle responds to the local gradient while the membrane shape changes with the evolving position of the vehicle, mimicking the interaction between matter and spacetime~\cite{wheeler1991geons}. The membrane shape is affected by three factors. Earth's gravity causes the membrane to sag with a parabolic profile due to the weight of the membrane itself. The tautness from boundary conditions such as the depth of the central depression of the membrane  competes with the sagging. Finally, the local deformation from the vehicle creates an additional dimple in  the membrane. We note that the 
sagging from the membrane's  weight and the deformation from the vehicle are analogous to the bending of background spacetime and the bending due to the moving object respectively  in the context of GR.

Our terrain consists of a spandex membrane fixed to a circular frame with a diameter of $2.4$ m.  The central depth of the membrane is controlled by an actuator that adjusts the height of the membrane center. The cylindrical chassis of the vehicle is 3D printed and its instantaneous position is tracked by an overhead camera systems (Optitrack).

% Further, we will compare our results to a standard sphere rolling with details shown later.

\begin{figure}[ht]
  \centering
  \includegraphics[width=0.65\textwidth]{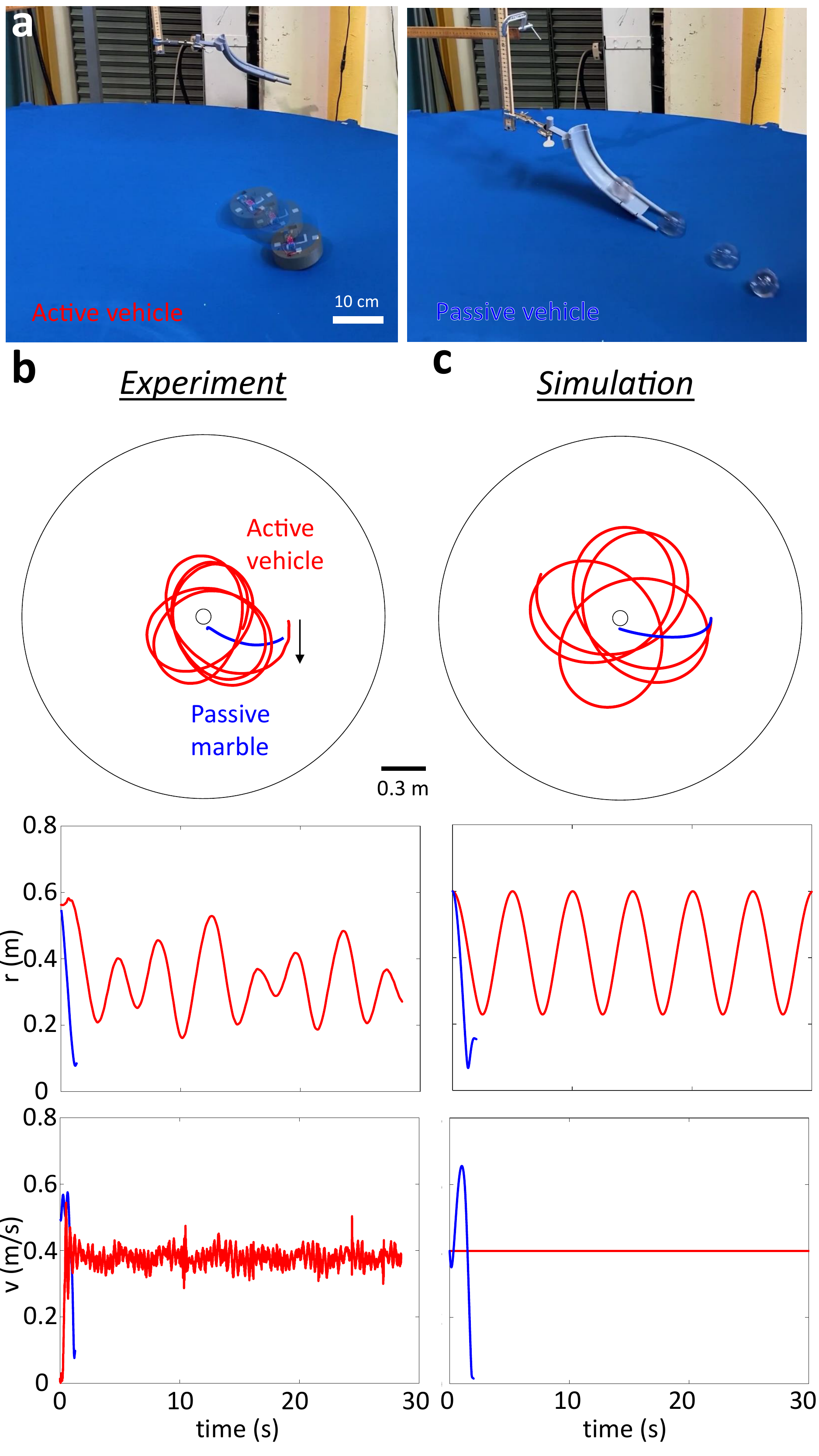}
  \caption{\textbf{Trajectories of passive and active objects on an elastic membrane.} (a) Sample perspective views of an active vehicle and a passive marble moving on a Spandex membrane. The time interval between two consecutive snapshots is $0.17$ s. (b) The experimental trajectories, radius evolution, and speed evolution of the active (red) and the passive (blue) objects with the same mass ($150$ g) started from the same initial position and velocity on the same membrane. See the \href{https://www.dropbox.com/s/zw9o9d8d0z8ckgl/GR_SciRep_movie.mp4?dl=0}{supplementary movie} for videos. (c) The simulation counterparts of (b).}
  %\vspace{-5mm}
  \label{fig:activePassiveExpt}
\end{figure}
% Remove the ramp and central depression on left figure. Add robot and its inset closeup. Add a marble on the right. An inset of the marble

We first compare the trajectories of the active vehicle with those of a passive marble having the same mass as the vehicle. We released the vehicle by placing it on the membrane after being turned on, and released the marble by placing it at the start of a guiding track. The velocity of the vehicle and marble were kept the same upon release by adjusting the voltage on the motor and the releasing height on the guiding track (Fig.\ref{fig:activePassiveExpt}a) respectively.  The trajectories collected from experiments showed that the active vehicle produced much more persistent trajectories (Fig.\ref{fig:activePassiveExpt}c) than the passive marble, which barely completed one  revolution (Fig.\ref{fig:activePassiveExpt}b).

\begin{figure}[ht!]
  \centering
  \includegraphics[width=0.75\textwidth]{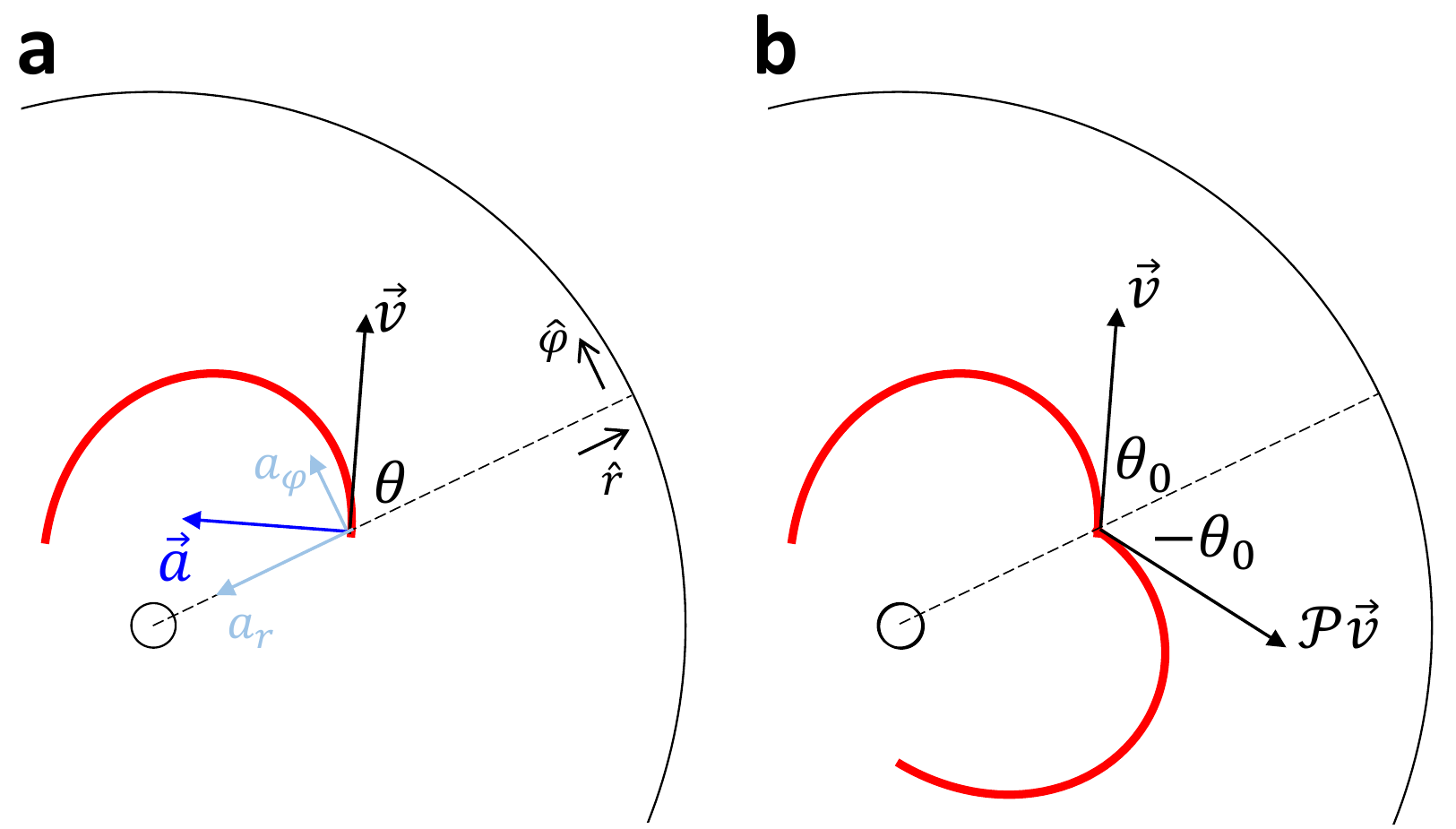}
  \caption{\textbf{Dynamics of the active vehicle.} (a) The acceleration of an active vehicle is perpendicular to its velocity $\vec{v}$. (b) A non-chiral vehicle with a mirror-reflected initial velocity $\mathcal{P}\vec{v}$ will produce a mirror-reflected trajectory.}
  %\vspace{-5mm}
  \label{fig:dynamics}
\end{figure}

To understand these orbits, we follow the models in \cite{fieldMediated,middleton2014circular}. While a passive marble dissipates energy as $\vec{a}\cdot\vec{v}<0$ (Fig.\ref{fig:activePassive}), an active object can conserve its speed when the driving force dynamically balances with the friction and exactly makes $\vec{a}\cdot\vec{v}=0$ (Fig.\ref{fig:dynamics}a). Therefore, the acceleration for a constant-speed motion can be written as
\begin{eqnarray}
 \frac{a_{\varphi}}{r} &=&\ddot{\varphi} + \frac{2\,\dot r \,\dot\varphi}{r} =  \frac{a}{r}\cos{\theta}\label{eq:reformulate1} \\
a_r &=&      \ddot{r} - \frac{r \,\dot{\varphi}^2}{\Psi^2} + \frac{\Psi'}{\Psi}\dot r^2 = -\frac{a}{\Psi}\sin{\theta}\label{eq:reformulate2},
\end{eqnarray}

where $\theta$ is the heading angle between the radial direction and the velocity on an isotropic circular membrane.

Though the speed is constant, the change of the velocity (the scalar acceleration $a$) depends on the local slope $\gamma$ (Fig.\ref{fig:activePassive}). Since $\gamma$ varies with radius (position) $r$, $a$ is also a function of $r$. Additionally, $a$ should also depend on velocity in general. However, given that the velocity has constant magnitude as the speed is constant, this dependence is reduced to one degree of freedom. For our convenience, we chose the direction of the velocity, $\theta$. If we consider an active object without chiral bias such that its trajectory has a mirror symmetry, the dependence of $a$ (thus $a_{\varphi}$) on $\theta$ should be anti-symmetric about $\theta=0$, as otherwise the clockwise ($\theta(t=0)=\theta_0$) and counterclockwise ($\theta(t=0)=-\theta_0$) trajectories (Fig.\ref{fig:dynamics}b) will not be mirror reflections with each other. A first-order approximation with this symmetry could be $a\propto k(r)\sin{\theta}$ where the $k(r)$ is the radial dependence due to the local slope $\gamma(r)$ that changes with radius. One could imagine $k$ increases with the local slope $\gamma$. The detailed  relation between $k$ and $\gamma$ depends on the mechanical structure of the active object, but one can  always Taylor expand this dependence. For preliminary study, here we assume linear dependence $k=C\gamma$.

While an active object follows equations Eqs.~\ref{eq:reformulate1}, \ref{eq:reformulate2}, a passive marble rolling on the membrane without slipping has a Lagrangian \cite{middleton2014circular}
\begin{eqnarray}
\mathcal{L}=\frac{7}{10}m\left((1+z'(r)^2)\dot{r}^2+r^2\dot{\varphi}^2\right)-mgz(r).
\end{eqnarray}
% XX WHAT MEANS EFFECTIVE COULOMB ROLLING FRICTION?
We consider the Coulomb rolling friction $-\mu mg \hat{\boldsymbol{v}}$ for a more realistic model. We neglect the higher-order effect ($O(\gamma^2)$) from the slope on the normal force for simplicity. Plugging the corresponding dissipation function \cite{goldstein2002} $D=\mu mg v$ where $v=\sqrt{\dot{r}^2+r^2\dot{\varphi}^2}$ into the Euler-Lagrange equation $\frac{d}{dt}(\frac{\partial\mathcal{L}}{\partial\dot{q}})-\frac{\partial\mathcal{L}}{\partial q}=-\frac{\partial D}{\partial\dot{q}}$ where $q$ is $r$ or $\varphi$, we arrive at
\begin{eqnarray}
(1+z'^2)\ddot{r}+z'z''\dot{r}^2-r\dot{\varphi}^2+\frac{5}{7}gz'&=&-\frac{5}{7}\mu g\frac{\dot{r}}{v}\label{eq:passive1}\\
r^2\ddot{\varphi}+2r\dot{r}\dot{\varphi}&=&-\frac{5}{7}\mu g\frac{r^2\dot{\varphi}}{v}\label{eq:passive2}
\end{eqnarray}
%\begin{eqnarray}
%(1+z'^2)\ddot{r}+z'z''\dot{r}^2-r\dot{\varphi}^2+\frac{5}{7}gz'=0\\
%\dot{\varphi}=\frac{5\ell}{7r^2}
%\end{eqnarray}

The left-hand sides of the above equations are the same as the dynamical equations in \cite{middleton2016elliptical} while the right hand sides correspond to the friction force.

Integration of the above models for the active vehicle and passive marble (Fig.\ref{fig:activePassiveExpt}c) shows qualitative agreement with the experiments. Fig.\ref{fig:activePassiveExpt}c shows the integration of the active dynamics Eqs.~\ref{eq:reformulate1}, \ref{eq:reformulate2} and the passive dynamics Eqs.~\ref{eq:passive1}, \ref{eq:passive2} on  the same simulated membrane with parameters measured from experiments. The active and passive objects started with  the same position and velocity. The physical parameters are measured from experiments. The acceleration dependence on radius $k$ for the active vehicle uses $k=C\gamma=C\partial_r z$ where $z(r)$ is measured from the height of the static vehicle placed at different radii $r$. The proportionality $C$ uses the ratio between acceleration and the gradient $\partial_r z$ at the radius close to the edge of the elastic membrane. We probe the friction coefficient for the passive marble by measuring the dissipation of mechanical energy in a designed experiment (see \hyperref[method:friction]{Sec.A} of the Methods section).\\

\section{Finding the spacetime corresponding to the orbits}

A functional understanding of the orbital features (e.g., period, precession rate) would facilitate the creation of an intelligently programmed vehicle. One tool we can use to obtain such insights is the spacetime metric that describes these orbits. Any similarities between the experimentally inferred metric and known metrics could prove useful in understanding how the orbital features depend on system parameters.

% \Gongjie{perhaps you mean: ``to make the analogy with GR in the weak field limit''?}\Shengkai{Yes. Have now added `in the weak field limit.'}
The orbital dynamics we wish to map could be described by a diversity of metrics. We propose a simple but general metric that is analogous to GR in the weak-field limit and encodes the axi-symmetry of the system. Our metric has the form
\begin{equation}
  ds^2 = -\alpha^2 dt^2 +\Phi^2(\Psi^2 dr^2 + r^2 d\varphi^2)
  \label{eq:metrics}
\end{equation}
with $\alpha = \alpha(r)$, $\Phi = \Phi(r), \Psi^2=1+z'^2$. Here, the elements of the metric $g_{\alpha\beta}$ are zero except $g_{tt} = -\alpha^2$, $g_{rr} = \Phi^2\Psi^2$ and $g_{\varphi\varphi} = \Phi^2r^2$. Inserting  $g_{\alpha\beta}$ into the Christoffel symbols $\Gamma^{\mu}_{\alpha\beta}$ in the geodesic equations $\ringring{x}^{\mu}+\Gamma^{\mu}_{\alpha\beta}\mathring{x}_{\alpha}\mathring{x}_{\beta}=0$, we arrive at
\begin{eqnarray}
&&\ringring{t} + \frac{(\alpha^2)'}{\alpha^2}\mathring{t}\mathring{r}= \frac{1}{\alpha^2}\left(\alpha^2\mathring{t}\right)^\circ = 0 \label{eq:geo1} \\
&&\ringring{\varphi} + \frac{(\Phi^2\,r^2)'}{\Phi^2\,r^2}\mathring{\varphi}\mathring{r}= \frac{1}{\Phi^2\,r^2}\left(\Phi^2\,r^2\mathring{\varphi}\right)^\circ =0 \label{eq:geo2}\\
&&\ringring{r} + \frac{(\alpha^2)'}{2\Phi^2\Psi^2}\mathring{t}^2+
\frac{(\Phi^2\Psi^2)'}{2\Phi^2\Psi^2}\mathring{r}^2 -\frac{(\Phi^2r^2)'}{2\Phi^2\Psi^2}\mathring{\varphi}^2 = 0\label{eq:geo3}
\end{eqnarray}
with $\lambda$ as an affine parameter and $\mathring{q}=dq/d\lambda,\ringring{q}=d^2q/d\lambda^2$. From Eqs.~\ref{eq:geo1},\ref{eq:geo2}, 
we have that
\begin{eqnarray}
\alpha^2\mathring{t} = E = \mathrm{constant}, \label{eq:E}\\
\Phi^2 r^2\mathring{\varphi} = L = \mathrm{constant}.\label{eq:L}
\end{eqnarray}
Both are consequences of conservation. Eq. \ref{eq:E} describes the conservation of energy while Eq. \ref{eq:L} describes the conservation of angular momentum.

Using  $\mathring{q}=(dq/dt)(dt/d\lambda)=\mathring{t}\dot{q}$ (see \hyperref[method:conversion]{Sec.B} of the Methods section for details), the geodesic equations can be rewritten as
\begin{eqnarray}
\ddot{\varphi} + \frac{2\dot r \dot\varphi}{r} &=&  \left[\frac{(\alpha^2)'}{\alpha^2} - \frac{(\Phi^2)'}{\Phi^2}\right]\dot r\,\dot\varphi\label{eq:GGeo1} \\
\ddot{r} - \frac{r \dot{\varphi}^2}{\Psi^2} + \frac{\Psi'}{\Psi}\dot r^2&=&  \left[\frac{(\alpha^2)'}{\alpha^2} - \frac{(\Phi^2)'}{\Phi^2} \right]\dot r^2\nonumber\\
&+& \frac{1}{2\,\Phi^2\Psi^2}\left[ (\Phi^2)'v^2-(\alpha^2)'\right]\ \label{eq:GGeo2}
\end{eqnarray}
where primes denote differentiation with respect to $r$.

Notice that the left-hand side of Eqs.~\ref{eq:GGeo1},\ref{eq:GGeo2} are the components of the acceleration, $a_{\varphi}$ and $a_r$ respectively, from   Eqs.~\ref{eq:reformulate1},\ref{eq:reformulate2}. When we substitute  $\cos{\theta}=\dot{r}/v,\sin{\theta}=r\dot{\varphi}/v$ and $a=k\sin{\theta}$ into Eqs.~\ref{eq:reformulate1},\ref{eq:reformulate2}, we find 
\begin{eqnarray}
\ddot{\varphi} + \frac{2\,\dot r \,\dot\varphi}{r} &=&  \frac{k}{v^2}\dot{r}\dot{\varphi}\label{eq:match1}\\
\ddot{r} - \frac{r \,\dot{\varphi}^2}{\Psi^2} + \frac{\Psi'}{\Psi}\dot r^2 &=& -\frac{k}{\Psi}\frac{r^2\dot{\varphi}^2}{v^2}.\label{eq:match2}
\end{eqnarray}

Thus, comparing the right-hand  sides of Eqs.~\ref{eq:GGeo1},\ref{eq:GGeo2} and Eqs.~\ref{eq:reformulate1},\ref{eq:reformulate2} and noticing that  $\dot{r}^2+r^2\dot{\varphi}^2=v^2$ in Eq.~\ref{eq:match2}, we obtain  the following relationships between the metric functions $\alpha$ and $\Phi$ in terms of the speed of the vehicle and $k$.
\begin{eqnarray}
  \frac{(\alpha^2)'}{\alpha^2} &=& \frac{k\Psi}{v^2}\left[\frac{\Phi^2v^2}{\alpha^2-\Phi^2v^2}\right] \label{eq:k1}\\
    \frac{(\Phi^2)'}{\Phi^2} &=& \frac{k\Psi}{v^2}\left[\frac{2\Phi^2v^2-\alpha^2}{\alpha^2-\Phi^2v^2}\right]\,.\label{eq:k2}
\end{eqnarray}

Integration of the above equations yields
\begin{eqnarray}
\alpha^2&=&-\frac{1}{C_1 v^2} + C_2 \cdot e^{-K/v^2}\label{eq:curvature1}\\
\Phi^2&=& \frac{\alpha^2}{v^2}+C_1 (\alpha^2)^2
\label{eq:curvature2}
\end{eqnarray}
where $K=K(r) \equiv \int_0^r k(s) \Psi(s) ds$. To determine the constants, we make use of the normalization condition and the fact that the metric should be flat at $k\rightarrow 0$.

The metric (Eq.~\ref{eq:metrics}) gives us the  normalization condition $-1 = -\alpha^2 \mathring{t}^2 + \Phi^2(\Psi^2\mathring{r}^2 + r^2 \mathring{\varphi}^2)$. To exploit this condition, we must  eliminate the $d/d\lambda$ in $\mathring{r}$ like Eqs.~\ref{eq:E}, \ref{eq:L}. Using $\mathring{q}=(dq/dt)(dt/d\lambda)=\mathring{t}\dot{q}$, Eqs.~\ref{eq:E}, \ref{eq:L}, and the fact that $v^2=r^2\dot{\varphi}^2+\dot{r}^2$, we have
\begin{eqnarray}
  \mathring{r}^2 &=& \left(\frac{E}{\alpha^2}\dot{r}\right)^2\nonumber\\
  &=& \frac{E^2}{(\alpha^2)^2}\frac{1}{\Psi^2}(v^2-r^2\dot{\varphi}^2)\nonumber\\
  &=&\frac{E^2}{(\alpha^2)^2}\frac{1}{\Psi^2}\left[v^2-r^2\left(\frac{\alpha^2}{E}\mathring{\varphi}\right)^2\right]\nonumber\\
  &=&\frac{E^2}{(\alpha^2)^2}\frac{1}{\Psi^2}\left[v^2-r^2\left(\frac{\alpha^2}{E}\frac{L}{\Phi^2 r^2}\right)^2\right].
\end{eqnarray}
Plug the $\mathring{t},\mathring{r},\mathring{\varphi}$ derived above into the normalization condition, we now have

\begin{eqnarray}
-1&=&-\frac{E^2}{\alpha^2} + \frac{\Phi^2 E^2 v^2}{(\alpha^2)^2}.
\end{eqnarray}

Plugging  in the $\alpha^2$ and $\Phi^2$ derived earlier (Eqs.~\ref{eq:curvature1},\ref{eq:curvature2}), we have $-\frac{1}{E^2}=C_1 v^2$
and therefore
\begin{eqnarray}
C_1=-\frac{1}{v^2 E^2}\label{eq:coeff_cond1}
\end{eqnarray}

Now, as promised earlier, we further determine the constants by making the metric flat when $k=0$.  In fact, $k(r)=0$ indicates $K(r)=\int_{s=0}^r k(s)\Psi(s) = 0$. We set the lower limit of the integral to zero, without loss of generality, since otherwise the arbitrary constant will be absorbed by $C_2$. This limit reduces the metric to $\alpha_0^2 =-\frac{1}{C_1 v^2} + C_2$ and therefore $\Phi_0^2= \frac{\alpha_0^2}{v^2}+C_1 (\alpha_0^2)^2$
where $\alpha_0\equiv\lim_{k\rightarrow 0}\alpha$ and $\Phi_0\equiv\lim_{k\rightarrow 0} \Phi$. To satisfy the flatness that $\alpha_0=\Phi_0$, we need $\alpha_0^2=\frac{\alpha_0^2}{v^2}+C_1 (\alpha_0^2)^2$. By using $\alpha_0^2=1/C_1 v^2 +C_2$ again, we have
\begin{eqnarray}
  C_1C_2&=&1.\label{eq:coeff_cond2}
\end{eqnarray}

The conditions in Eqs.~\ref{eq:coeff_cond1},\ref{eq:coeff_cond2} settle the previously undetermined coefficients in the metric (Eqs.~\ref{eq:curvature1},\ref{eq:curvature2}). We finally arrive at
\begin{eqnarray}
\alpha^2 &=& E^2(1-v^2 e^{-K/v^2})\label{eq:alpha1}\\
\Phi^2&=&E^2 e^{-K/v^2} (1-v^2 e^{-K/v^2}).\label{eq:Phi1}
\end{eqnarray}

The constants $E$ and $L$ required to fix the metric have actual physical meanings. $E$ is the constant energy associated with the fact that the metric is time-independent. $L$ is the constant angular momentum associated with the metric's $\varphi$-symmetry.

Thus our formulation indeed reveals that the vehicle does not simply follow spatial geodesics of the membrane but instead follows geodesics in an emergent spacetime (Eqs. \ref{eq:alpha1},\ref{eq:Phi1})  generated by the global curvature, the local curvature, the active dynamics, and the differential mechanism. The resultant dynamics can now be understood as those of a test particle in a new spacetime where the active feature of the real particle, such as a persistently controlled speed, generates a non-splittable effective spacetime for the test particle (i.e. $g_{tt}$ is not constant). In the language of the work by Price~\cite{price2016spatial}, the effects of curvature are now not restricted to space \cite{batlle2019exploring}. That is, in general, the metric function $g_{tt}$ could depend on both the coordinate time ($t$) as well as the spatial coordinates. For a static metric (i.e., the metric functions are independent of time), the spacetime becomes splittable when $g_{tt}$ does not depend on the spatial coordinates. This leads to only spatial curvature. It was argued in \cite{price2016spatial} that the spatial curvature is different from the spacetime curvature as it is devoid of gravity, i.e., a free particle initially at rest will remain at rest.

The essential contribution from the  active drive is the persistent response to the local curvature, here enabled by the controlled constant speed unseen in passive systems. In fact, when the response of the turning to the local slope vanishes at the limit $v\rightarrow \infty$ such that $\alpha^2 = \Phi^2 = E^2(1-v^2)$, the metric Eq.~\ref{eq:metrics} with components Eqs.~\ref{eq:alpha1},\ref{eq:Phi1} reduces to a splittable (and flat) spacetime Eq.~\ref{eq:spaGeod}. On the other hand, when $v$ is finite and controllable, the active locomotion provides more flexibility and programmability in constructing the desired spacetime in GR. For instance, programming an active agent with acceleration magnitude $k$ decreasing with radius $r$ makes an orbit precess in the same direction as the orbit while a $k$ increasing with $r$ makes an orbit precess in the opposite direction as the orbit \cite{fieldMediated}. Such flexibility and programmability are more challenging in passive and dissipative agents studied in the previous works\cite{middleton2016elliptical,middleton2014circular}. \\

\section{Programming an  arbitrary spacetime with a  speed-varying robot}

The metric Eqs.~\ref{eq:alpha1},\ref{eq:Phi1} has shown us how the parameters of the system change the spacetime and thus the orbit. Now we see how we can solve  the inverse problem of programming  the desired spacetime using  the system parameters (e.g., $k(r)$ and $v(r)$). %\Gongjie{(e.g., $k(r)$ and $v(r)$)}\Shengkai{Added.}.

In metric Eqs.~\ref{eq:alpha1},\ref{eq:Phi1}, we can tune the speed and membrane elasticity to change the spacetime of the orbits. However, here the spatial and radial metric are not yet  completely disentangled yet. To have two degrees of freedom such that we can indeed program the spacetime arbitrarily, one could introduce another degree of freedom. For instance, if we allow the speed $v$ to vary with the radius $r$ (physical instantiation could be achieved by inferring the radius from the instantaneous tilting angle $\gamma$), Eqs. \ref{eq:k1}, \ref{eq:k2} with $\Psi^2\approx 1$ give the requirement for the  mapping as

\begin{eqnarray}
    \frac{(\alpha^2)'}{\alpha^2}-\frac{(\Phi^2)'}{\Phi^2}&=&\frac{v'}{v}+\frac{k}{v^2}\label{eq:prog1}\\
    \frac{(\Phi^2)'v^2-(\alpha^2)'}{2\Phi^2}&=&-k.\label{eq:prog2}
\end{eqnarray}

These two equations above give us the recipe to create a desired spacetime by changing the speed of the vehicle with radius. For a desired metric with spatial curvature $\Phi^2(r)$ and temporal curvature $\alpha^2(r)$, we can solve for the required membrane elasticity and object speed by plugging in the curvatures into these two equations. The solution (see \hyperref[method:programming]{Sec.C} of the Methods section for details) is

\begin{eqnarray}
v(r)^2=\left(\int_{r_1}^r f(r')\cdot\frac{(\alpha^2)'(r')}{\Phi^2(r')}dr'\right)/f(r)
\end{eqnarray}

where

\begin{eqnarray}
f(r)=-e^{\int_{r_1}^r -2\frac{(\alpha^2)'(r')}{\alpha^2(r')}+\frac{(\Phi^2)'(r')}{\Phi^2(r')}dr'}.
\end{eqnarray}

\begin{figure}[ht]
  \centering
  \includegraphics[width=0.58\textwidth]{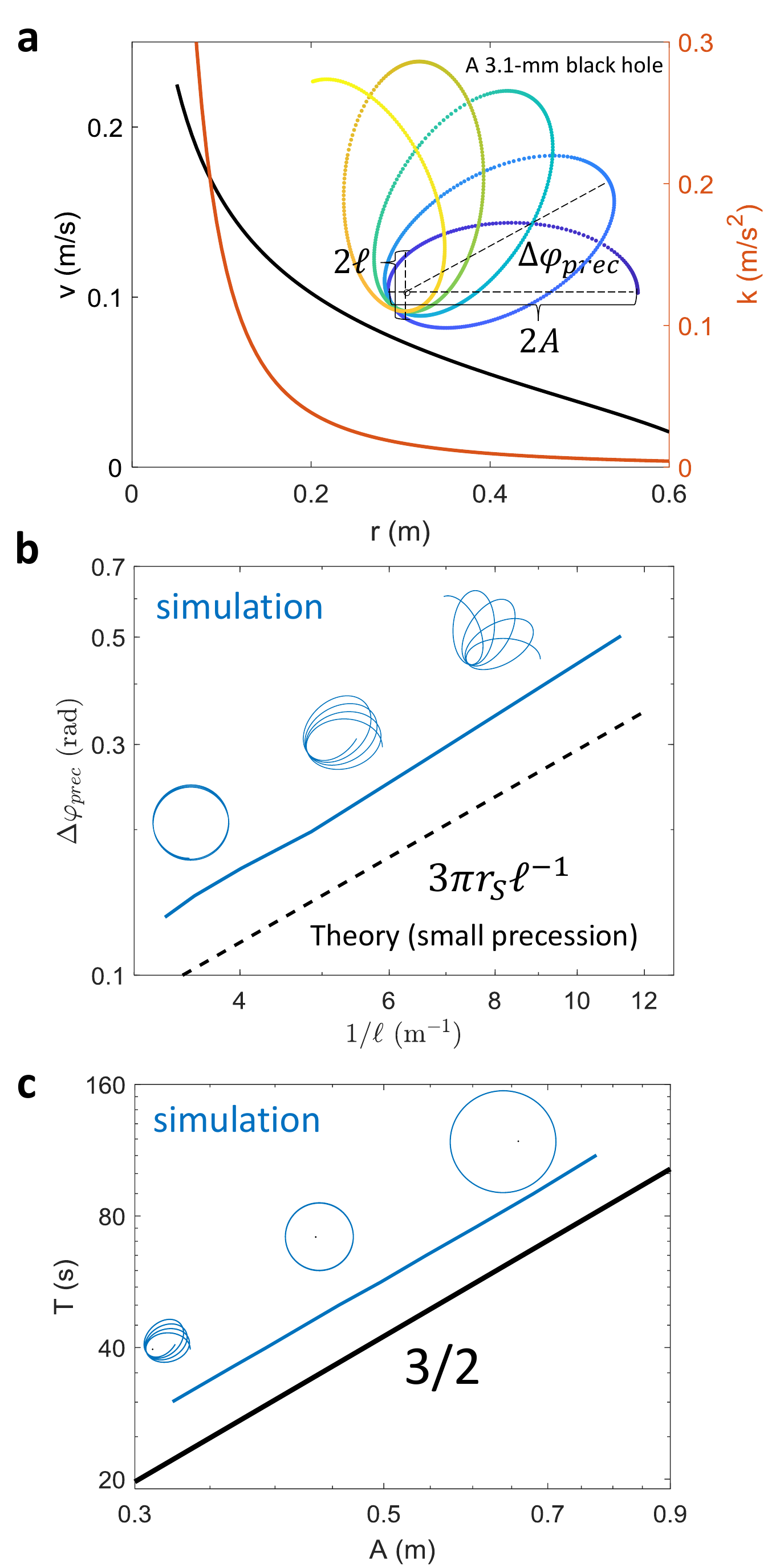}
  \caption{\textbf{Creating orbits in Schwarzschild spacetime  with a  speed varying particle.} (a) The speed and membrane elasticity's dependence on radius to create a Schwarzschild blackhole with $r_s=3.1$ mm. The inset shows a precessing orbit with $A=0.3$ m using this prescription. (b) Precession angle $|\Delta\varphi_{\mathrm{prec}}|$ as a function of inverse latus rectum. (c) The relation between the orbital period, $T$, and the semi-major-axis, $A$, follows Kepler's third law as $T\propto A^{3/2}$. Insets in (b) and (c) show the trajectories around the data points. See the \href{https://www.dropbox.com/s/zw9o9d8d0z8ckgl/GR_SciRep_movie.mp4?dl=0}{supplementary movie} for continuous evolution of the orbits.}
  %\vspace{-5mm}
  \label{fig:schwarzschild}
\end{figure}

For instance, if we plug in the Schwarzschild metric in isotropic coordinates where  $\alpha^2(r)=1-r_s/r,\Phi^2(r)=(1-r_s/r)^{-1}$, we arrive at the required  membrane elasticity $k(r)$ and active object speed $v(r)$ as shown in Fig.\ref{fig:schwarzschild}a. Analytically,

\begin{eqnarray}
v(r)^2&=&r_s\frac{(r-r_s)^2}{r^3}+C\left(\frac{r-r_s}{r}\right)^3\\
k(r)&=&\frac{r_s(r-r_s)(r+Cr+r_s-Cr_s)}{2r^4}
\end{eqnarray}

where

\begin{eqnarray}
C=\frac{v_0^2r_0^3}{(r_0-r_s)^3}-\frac{r_s}{r_0-r_s}.
\end{eqnarray}

Here, $v_0$ is the vehicle speed at $r_0$ as the boundary condition. For instance,  one can use the inner radius as $r_0$.

Simulations using this prescription show features of Schwarzschild orbits such as the linear dependence of the precession angle in terms of the inverse latus rectum. For Schwarzschild orbits with small precession, the precession angle increases with the inverse latus rectum as $\Delta\varphi_{\mathrm{prec}} = 6\pi G^2M/(c^2l)=3\pi r_s \ell^{-1}$ where $G$ is the gravitational constant, $M$ is the mass of the star, $c$ is the speed of light, and $\ell\equiv A~(1-e^2)$ is the latus rectum. We evaluate the semi-major-axis $A$ and the eccentricity $e$ using the minimum and maximum radii: $A=(r_{\mathrm{max}}+r_{\mathrm{min}})/2,~e=(r_{\mathrm{max}}-r_{\mathrm{min}})/(r_{\mathrm{max}}+r_{\mathrm{min}})$. Fig.~\ref{fig:schwarzschild}b shows the precession angle $\Delta\varphi_{\mathrm{prec}}$ as a function of the inverse of the latus rectum, $\ell^{-1}$, from simulations given $v_0=v(r_0=0.05\mathrm{m})=0.225~\mathrm{m/s}$ and $r_s=0.0031$ m. The curve qualitatively follows the linear relationship, with small deviation from the theory due to the large precession angle. By changing $(r_0,v_0)$, we can obtain  larger angular momenta and thus larger orbits around the same blackhole. These orbits show a relation between period $T$ and semi-major-axis $A$ that follows Kepler's third law (Fig.\ref{fig:schwarzschild}c).% \Gongjie{Perhaps we could list some caveats on why the simulation doesn't agree with the theory?} 

To achieve this relation  in experiments, a vehicle must actively vary its speed with radius and a membrane must have a radially-dependent elastic modulus. We can  attach a tilt sensor to infer the radius and change the speed accordingly. To program the membrane with radially varying profile $k(r)=Cg|\partial_r z|$, here we consider a membrane with linear elasticity following the Poisson Equation $\nabla\cdot E \nabla z = P$ where $P$ is the unit load from the membrane gravity. One possible way to obtain the desired $k(r)$ is to create an elastic material with a radially varying thickness $P=P(r)$. Another option is to fabricate a membrane with a radially varying modulus $E=E(r)$.\\

\section{Programming a spacetime with a non-diagonal metric}
\begin{figure}[ht]
  \centering
  \includegraphics[width=0.9\textwidth]{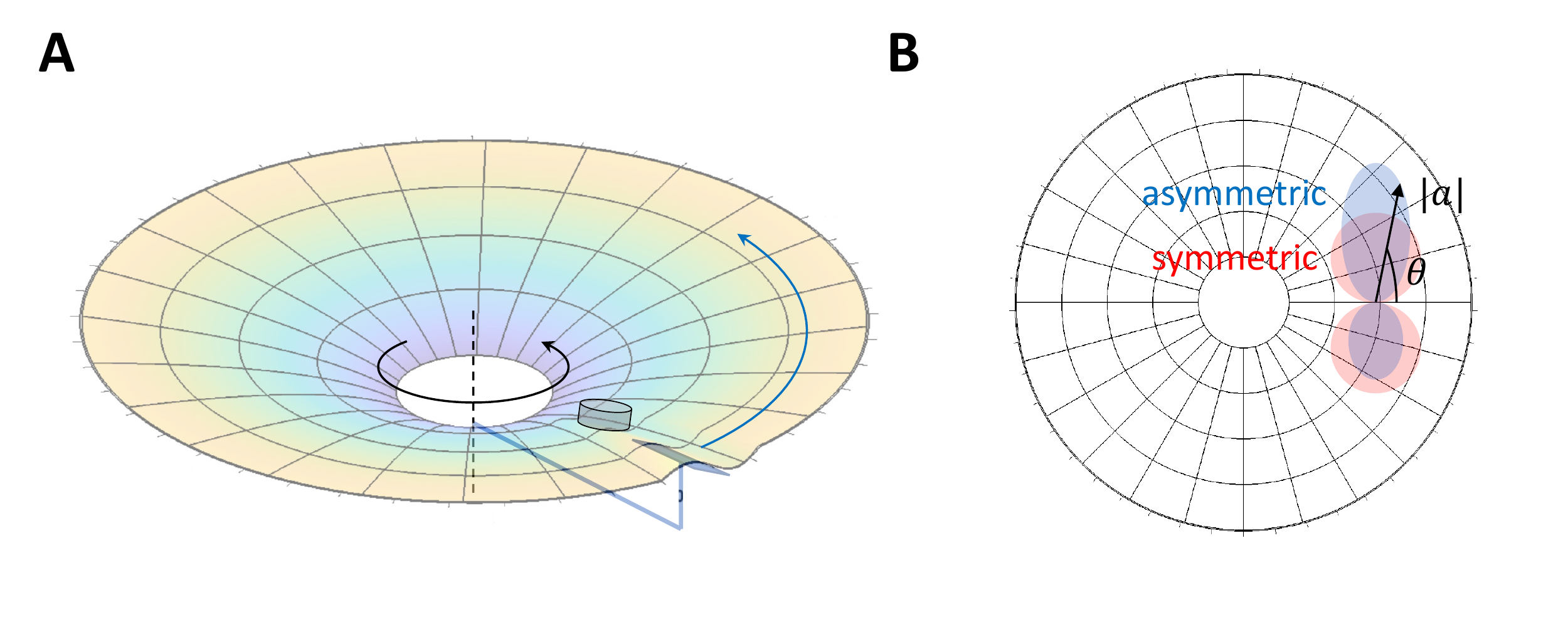}
  \caption{\textbf{Proposed apparatus to create a spacetime with a non-diagonal metric.} A controlled deformation (e.g., a tilted slab) rotating about the central axis shown in (A) could plausibly induce an asymmetric dependence on heading $\theta$ for the acceleration magnitude in blue instead of the symmetric counterpart in red. The spacetime metric resulting from this proposed setup would have a non-diagonal component $g_{t\phi}$.}
  %\vspace{-5mm}
  \label{fig:asymmetry}
\end{figure}

In principle, one can even create orbits from a non-diagonal metric by breaking symmetries. Here we show that breaking the axis symmetry can create a spacetime with nonzero $g_{t\phi}$, which is essential in perhaps the most well-known non-diagonal metric, the Kerr metric for a rotating blackhole. Experimentally, this could be done by adding tangential perturbation to the substrate. See Fig.\ref{fig:asymmetry} for an illustration.

% For instance, the Kerr metric at the equator has $g_{tt}=-(1-\frac{2Mr}{\Sigma}),g_{rr}=\Sigma/\Delta,g_{t\phi}=-\frac{Mr}{\Sigma}a,g_{\phi\phi}=r^2+a^2+\frac{2Mra^2}{\Sigma},g^{tt}=-\frac{1}{\Delta}(r^2+a^2+\frac{2Mra^2}{\Sigma}),g^{rr}=\Delta/\Sigma,g^{t\phi}=-\frac{2Mr}{\Sigma\Delta}a,g^{\phi\phi}=\frac{\Delta-a^2}{\Sigma\Delta}$ where $\Sigma=r^2,\Delta=r^2-2Mr+a^2$ with $M$ and $Ma$ being the mass and angular momentum of the blackhole.

Let us consider a $(2+1)$D metric $g_{\mu\nu}$ where the only off-diagonal term is the $g_{t\phi}$. Following the same methodology we used for the Schwarzschild metric, the geodesic equation is now

\begin{eqnarray}
\ringring{t}&=&-\Gamma^t_{tr}\mathring{t}\mathring{r}{\color{red}-\Gamma^t_{r\phi}\mathring{r}\mathring{\phi}}\\
\ringring{r}&=&-\Gamma^r_{tt}\mathring{t}^2{\color{red}-\Gamma^r_{t\phi}\mathring{t}\mathring{\phi}}-\Gamma^r_{rr}\mathring{r}^2-\Gamma^r_{\phi\phi}\mathring{\phi}^2\\
\ringring{\phi}&=&{\color{red}-\Gamma^{\phi}_{tr}\mathring{t}\mathring{r}}-\Gamma^{\phi}_{r\phi}\mathring{r}\mathring{\phi}
\end{eqnarray}

where the red terms are from the off-diagonal components in addition to the diagonal metric (for instance, Schwarzschild) we worked on in the previous section. See \hyperref[method:offDiagonal]{Sec.D} of the Methods section for the technical details of this section.

The conserved quantities are now generalized to be $Q_t=P_{tt}(r)\mathring{t}+P_{t\phi}\mathring{\phi}$ and $Q_{\phi}=P_{\phi t}(r)\mathring{t} +P_{\phi\phi}\mathring{\phi}$ where
\begin{eqnarray}
P^{-1}P'=
\begin{pmatrix}
\Gamma_{tr}^t&{\color{red}\Gamma_{r\phi}^t}\\ {\color{red}\Gamma_{tr}^{\phi}}&\Gamma_{r\phi}^{\phi}
\end{pmatrix} \text{with } P\equiv
\begin{pmatrix}
    P_{tt} & P_{t\phi}\\
    P_{\phi t} & P_{\phi\phi}
\end{pmatrix}.
\end{eqnarray}

While $\Gamma_{r\phi}^t$ and $\Gamma_{tr}^{\phi}$ are now nonzero such that $P$ is not simply $\text{diag}\{\alpha^2(r),\Phi^2(r)r^2\}$ as we have in the diagonal case, the conserved quantities $(Q_t,Q_{\phi})^T=P(\mathring{t},\mathring{\phi})^T$ still give us $(\mathring{t},\mathring{\phi})^T=P^{-1}(Q_t,Q_{\phi})^T\equiv (f(r),g(r))^T$ as functions of $r$. From this, we obtain the equations of motion with respect to time $t$ instead of affine parameter $\lambda$ as
\begin{eqnarray}
\ddot{\phi}+\frac{2\dot{r}\dot{\phi}}{r}&=&c_{r\phi}\dot{r}\dot{\phi}-\Gamma_{tr}^{\phi}\dot{r}\label{eq:Kerr1}\\
\ddot{r}-r\dot{\phi}^2&=&c_{\phi\phi}\dot{\phi}^2-\Gamma_{t\phi}^r\dot{\phi}+c_0\label{eq:Kerr2}
\end{eqnarray}
where the $c$'s are all functions of $r$ such that $c_{\phi\phi}=(f'/f+\Gamma_{rr}^r)r^2-\Gamma_{\phi\phi}^r-r$, $c_0=-\Gamma_{tt}^r-(f'/f+\Gamma_{rr}^r)v^2$, $c_{r\phi}=-\Gamma_{r\phi}^{\phi}-f'/f+2/r$.

To accommodate the new terms with $\dot{\phi}$ and $\dot{r}$, one could use a rotating object \cite{drill} (for instance a tilted slab shown in Fig.\ref{fig:asymmetry}) to locally add an azimuthal perturbation $\delta$. This perturbation would break the symmetry such that the magnitude of acceleration $a=k (\sin{\theta} + \delta)$ has a bias in one chirality over the other since $a(\theta')\neq a(-\theta')$. Noting $\sin{\theta}=r\dot{\phi}/v$, the equation of motion of the vehicle on a substrate with broken axial symmetry is
\begin{eqnarray}
\ddot{\phi}+\frac{2\dot{r}\dot{\phi}}{r}&=&\frac{1}{r}\frac{\dot{r}}{v}k(\frac{r\dot{\phi}}{v}+\delta)\label{eq:offDiag1}\\
\ddot{r}-r\dot{\phi}^2&=&-\frac{r\dot{\phi}}{v}k(\frac{r\dot{\phi}}{v}+\delta)\label{eq:offDiag2}
\end{eqnarray}

$\delta$ is a function of $r$, which can be determined by matching the $r$-dependent functions in Eqs.\ref{eq:Kerr1}-\ref{eq:Kerr2} and Eqs.\ref{eq:offDiag1}-\ref{eq:offDiag2} either analytically or numerically.
% https://www.roma1.infn.it/teongrav/onde19_20/kerr.pdf

\section{Discussion}

In this work, we demonstrated how the use of an active particle -- a wheeled robot -- moving on an elastic membrane can generate a system which can mimic the dynamics of bodies in arbitrary spacetime. Given the flexibility in constructing  and programming such robophysical devices, our system makes for an attractive target to push toward a mechanical analog GR system. While superficially our system resembles the educational tool used to motivate Einstein's view of spacetime curvature influencing matter trajectories \cite{white2002shape, middleton2014circular,middleton2016elliptical}, unlike such systems which are \textit{not} good analogs of GR, the activity allows the dynamics of the vehicle to be dictated by the curvature of ``spacetime", not just the curvature of space as in splittable spacetimes ($g_{tt}$ is constant)~\cite{price2016spatial}. As such, our system can be used as an experimental example of GR in upper-division physics courses to enhance students' hands-on understanding \cite{holmes2015teaching,smith2020direct} of orbits and curved spacetime. Thus we posit that mechanical analog ``robophysical'' \cite{aguilar2016review,AydinBookChap} systems can complement existing fluid \cite{unruh1981experimental,patrick2018black}, condensed matter \cite{steinhauer2016observation}, atomic, and optical \cite{zhu2018elastic,philbin2008fiber,belgiorno2010hawking} analog gravity systems \cite{Barcelo2011} given the ability to create infinite types of spacetimes. We might even generate analogies to wave-like systems \cite{couder2005dynamical,bush2015pilot,BushQuantum}; for example, one could increase the speed of the vehicle to be comparable to disturbance propagation (such that the membrane would follow the wave equation).  There has been much theoretical progress on the analogy between spacetime and solid mechanics \cite{tenev2018mechanics}. We posit mechanical devices such as the one we introduce in this work could help explore the theoretical proposal with the reverse methodology. For instance, we can design systems to monitor the elasticity of known spacetimes and probe interesting problems such as the occurrence of topological defects and plastic deformations in spacetime.

We advocate for the potential research applications of our robotic vehicle as an analog system. In astrophysics, we typically are limited to an observational understanding of phenomena borne out of an inability to perform physical experiments with customized parameters. Though simulation provides an alternative, limitations such as sub-grid physics and runtime practicality for high-resolution or multiscale dynamics \cite{heng2014nature}, path dependency and contingency of composite models \cite{ruphy2011limits}, and the nature of modeling approximation when carrying out explorative research \cite{gruner2020computer} are nontrivial obstacles that may hinder the validity and applicability of these simulations. Physical analog systems provide simple parameter variation, intellectual accessibility, and potentially offer ease of solution due to their classical nature \cite{Barcelo2011}. With the flexibility of parameters introduced in this paper,  the length and time scales not accessible in celestial systems are possible to access in their laboratory analogs. For instance, it is possible to probe physics around or even within the event horizon of a blackhole to probe theories \cite{maluf2016teleparallel} that are challenging to test in the original system. See \hyperref[method:friction]{Sec.E} of the Methods section for one such example of repulsive orbits inside the event horizon.

Beyond its role as a mechanical analog for GR, this framework could also provide a new perspective to understanding active matter undergoing field-mediated interactions \cite{wang2021emergent,fieldMediated}. For instance, the spacetime metric of the agents' motion can both guide our choice of parameter values to alter orbital features, like the precession sign, and influence our design of control schemes that accomplish tasks like helping multiple agents avoid mergers on the membrane \cite{fieldMediated}.

\section{Acknowledgments}
 We thank Lutian Zhao for the help on the differential equation. Funding for S.L. and D.I.G. was provided by the Army Research Office (ARO) grant no. W911NF-21-1-0033 and the MURI award no. W911NF19-1-0233; funding was also provided by Dunn Family Professorship. P.L. was supported by NSF Grants 2207780, 2114581, and 2114582. G.L. was supported by NASA Grants 80NSSC20K0641 and 80NSSC20K0522.

\section{Data availability}

All data generated or analysed during this study are included in this published article and its supplementary information files.

%\begin{appendices}
\section*{Methods}
\subsection*{\textbf{A. Probing the effective friction}}\label{method:friction}

\begin{figure}[ht]
  \centering
  \includegraphics[width=0.6\textwidth]{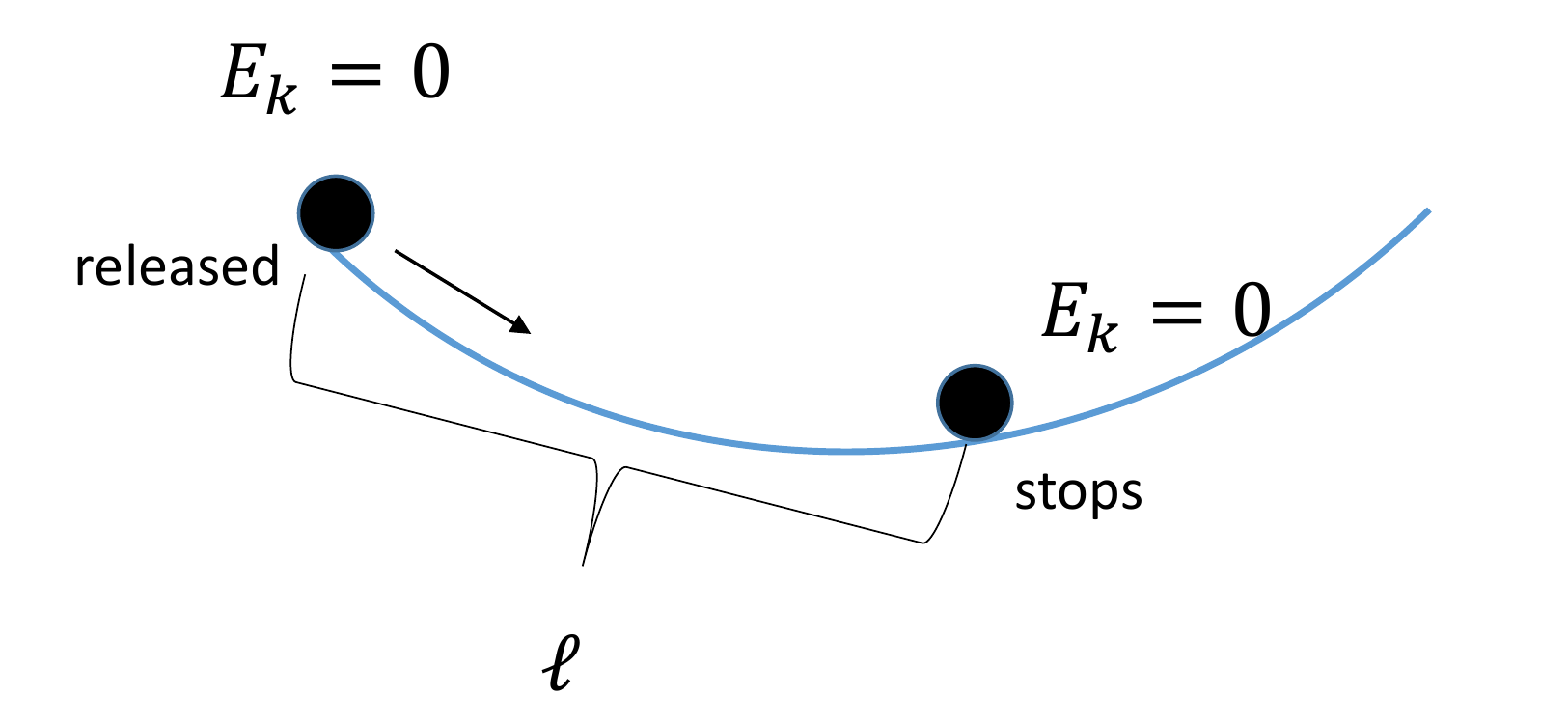}
  \caption{\textbf{An experiment to probe the effective friction.}}
  \vspace{-5mm}
  \label{fig:frictionMeasurement}
\end{figure}

We simplify the complicated rolling friction and membrane dissipation by employing an effective friction constant that absorbs all dissipative forces. We probe its magnitude by doing the following experiment.  We release the marble at the rim of the circular membrane with zero speed and thus zero kinetic energy (Fig.\ref{fig:frictionMeasurement}). The marble then rolls radially towards the center, passes through the center, and stops before it reaches the other end of the diameter due to the effective rolling friction. Absorbing the loss of mechanical energy into the dissipation from the effective rolling friction $f_{\mathrm{roll}}$ for a distance of $\ell$, we arrive at
\begin{eqnarray}
f_{\mathrm{roll}}\ell=mg\Delta h
\end{eqnarray}
The measurements from experiment that $\ell = 1.5$ m, $\Delta h = 0.1$ m give the effective friction coefficient $\mu = f_{\mathrm{roll}}/mg=\Delta h/\ell = 0.07 \sim 0.1$.

\newpage
\subsection*{\textbf{B. Converting derivatives}}
\label{method:conversion}

With the help of $\mathring{q}\equiv\frac{dq}{d\lambda} = \frac{dt}{d\lambda}\frac{dq}{dt}$ and $\alpha^2\mathring{t} = E, \Phi^2 r^2\mathring{\varphi} = L$ in Eqs.~\ref{eq:E},\ref{eq:L}, we have
\begin{eqnarray}
  \mathring{t} &=& \frac{E}{\alpha^2}\\
  \ringring{t} &=& \frac{dt}{d\lambda}\frac{d\mathring{t}}{dt}\nonumber\\
  &=&\frac{E}{\alpha^2} \frac{d}{dt}(\frac{E}{\alpha^2})\nonumber\\
  &=&-\frac{E^2 (\alpha^2)'}{(\alpha^2)^3} \dot{r}\\
  \mathring{r}&=&\frac{dt}{d\lambda}\frac{dr}{dt}=\frac{E}{\alpha^2}\dot{r}\\
  \ringring{r}&=&\frac{dt}{d\lambda}\frac{d\mathring{r}}{dt}\nonumber\\
  &=&\frac{E}{\alpha^2}\cdot\frac{d}{dt}(\frac{E}{\alpha^2}\dot{r})\nonumber\\
  &=&\frac{E^2}{(\alpha^2)^2}\cdot \left(-\frac{(\alpha^2)'}{\alpha^2}\dot{r}^2+\ddot{r}\right)\\
  \mathring{\varphi}&=&\frac{dt}{d\lambda}\frac{d\varphi}{dt}=\frac{E}{\alpha^2}\dot{\varphi}\\
  \ringring{\varphi}&=&\frac{dt}{d\lambda}\frac{d\mathring{\varphi}}{dt}\nonumber\\
  &=&\frac{E}{\alpha^2 } \frac{d}{dt}(\frac{E}{\alpha^2}\dot{\varphi})\nonumber\\
  &=&\frac{E^2}{(\alpha^2)^2}\cdot \left(-\frac{(\alpha^2)'}{\alpha^2}\dot{r}\dot{\varphi}+\ddot{\varphi}\right).
\end{eqnarray}

\quad\\

\subsection*{\textbf{C. Programming the metric}}\label{method:programming}

By eliminating the $k$ in Eqs.~\ref{eq:prog1},\ref{eq:prog2}, we get
\begin{eqnarray}
M V-V'=\frac{(\alpha^2)'}{\Phi^2}\label{eq:nExact}
\end{eqnarray}

where $M(r)=2(\alpha^2)'(r)/\alpha^2(r)-(\Phi^2)'(r)/\Phi^2(r)$ and $V(r)=v^2(r)$.

We can multiply a function $f(r)$ to both sides of Eq.~\ref{eq:nExact} to make the left-hand side exact. Noting $(fV)'=f'V+fV'$, we need $f'/f=-M$. Therefore,
\begin{eqnarray}
f(r)=-e^{\int_{r_1}^r -M(r')dr'}.
\end{eqnarray}

With this $f$, we now have $(fV)'=f~(\alpha^2)'/\Phi^2$. So,
\begin{eqnarray}
V(r)=\left(\int_{r_1}^r f(r')\cdot\frac{(\alpha^2)'(r')}{\Phi^2(r')}dr'\right)\cdot\frac{1}{f(r)}.
\end{eqnarray}

By plugging in the Schwarzschild metric in isotropic coordinates $\alpha^2(r)=1-r_s/r,\Phi^2(r)=(1-r_s/r)^{-1}$, we have
\begin{eqnarray}
f(r)&=&-e^{\int_{r_1}^r -\frac{3r_s}{r'(r'-r_s)}dr'}\nonumber\\
&=&-e^{(C_1+3\log{(r/(r-r_s))})}\nonumber\\
&=&-C_2\cdot\left(\frac{r}{r-r_s}\right)^3.
\end{eqnarray}
Therefore,
\begin{eqnarray}
V(r)&=&\left(\int_{r_1}^r C_2\cdot\left(\frac{r'}{r'-r_s}\right)^3\cdot\frac{(r'-r_s)r_s}{r'^3}dr'\right)/f(r)\nonumber\\
&=&\left(\int_{r_1}^r C_2\cdot \frac{r_s}{(r'-r_s)^2}dr'\right)/f(r)\nonumber\\
&=&(-C_2\frac{r_s}{r-r_s}+C_3)/(-C_2\cdot\left(\frac{r}{r-r_s}\right)^3)\nonumber\\
&=&r_s\frac{(r-r_s)^2}{r^3}+C\left(\frac{r-r_s}{r}\right)^3\\
k(r)&=&-\frac{(\Phi^2)'V-(\alpha^2)'}{2\Phi^2}\nonumber\\
&=&\frac{(r-r_s)r_s(r+Cr+r_s-Cr_s)}{2r^4}.
\end{eqnarray}

To program the active object physically, we want to prescribe the speed $v_0$ at a certain radius (say the inner radius $r_0$) so that $V(r_0)=v_0^2$, we need

\begin{eqnarray}
C=\frac{v_0^2r_0^3}{(r_0-r_s)^3}-\frac{r_s}{r_0-r_s}.
\end{eqnarray}

Further, a reasonable speed $v_c$ at a characteristic orbit size (say the circular orbit $r_c$) will limit the size of the Schwarzschild radius $r_s$ (the size of the blackhole) with $\frac{V(r_c;r_s)}{r_c}=k(r_c)$.

\subsection*{\textbf{D. Programming metric with off-diagonal}}\label{method:offDiagonal}
For metric with nonzero $g_{t\phi}$, the nonzero Christoffel symbols are
\begin{eqnarray}
{\Gamma^t_{tr}}&=&{\frac{1}{2}(g^{tt}g_{tt,r}+g^{t\phi}g_{\phi t,r})}\\
{\Gamma^t_{r\phi}}&=&{\frac{1}{2}(g^{tt}g_{t\phi,r}+g^{t\phi}g_{\phi\phi,r})}\\
{\Gamma^r_{tt}}&=&{-\frac{1}{2}g^{rr}g_{tt,r}}\\
{\Gamma^r_{t\phi}}&=&{-\frac{1}{2}g^{rr}g_{t\phi,r}}\\
\end{eqnarray}
\begin{eqnarray}
{\Gamma^r_{rr}}&=&{\frac{1}{2}g^{rr}g_{rr,r}}\\
{\Gamma^r_{\phi\phi}}&=&{-\frac{1}{2}g^{rr}g_{\phi\phi,r}}\\
{\Gamma^{\phi}_{tr}}&=&{\frac{1}{2}(g^{\phi\phi}g_{\phi t,r}+g^{\phi t}g_{tt,r})}\\
{\Gamma^{\phi}_{r\phi}}&=&{\frac{1}{2}(g^{\phi t}g_{t\phi,r}+g^{\phi\phi}g_{\phi\phi,r})}
\end{eqnarray}
The geodesic equation is therefore
\begin{eqnarray}
\ringring{t}&=&-\Gamma^t_{tr}\mathring{t}\mathring{r}-\Gamma^t_{r\phi}\mathring{r}\mathring{\phi}\\
\ringring{r}&=&-\Gamma^r_{tt}\mathring{t}^2-\Gamma^r_{t\phi}\mathring{t}\mathring{\phi}-\Gamma^r_{rr}\mathring{r}^2-\Gamma^r_{\phi\phi}\mathring{\phi}^2\label{eq:nonDiag_r}\\
\ringring{\phi}&=&-\Gamma^{\phi}_{tr}\mathring{t}\mathring{r}-\Gamma^{\phi}_{r\phi}\mathring{r}\mathring{\phi}\label{eq:nonDiag_phi}
\end{eqnarray}

Assume now the conserved quantities are $Q_t=P_{tt}\mathring{t}+P_{t\phi}\mathring{\phi},Q_{\phi}=P_{\phi t}\mathring{t}+P_{\phi\phi}\mathring{\phi}$, then
\begin{eqnarray}
\frac{dQ_t}{d\lambda}=P'_{tt}\mathring{r}\mathring{t}+P_{tt}\ringring{t}+P'_{t\phi}\mathring{r}\mathring{\phi}+P_{t\phi}\ringring{\phi}&=&0\\
\frac{dQ_{\phi}}{d\lambda}=P'_{\phi t}\mathring{r}\mathring{t}+P_{\phi t}\ringring{t}+P'_{\phi\phi}\mathring{r}\mathring{\phi}+P_{\phi\phi}\ringring{\phi}&=&0
\end{eqnarray}
which can be also written as
\begin{eqnarray}
P\begin{pmatrix}
\ringring{t}\\ \ringring{\phi}
\end{pmatrix}
=
-P'\begin{pmatrix}
\mathring{r}\mathring{t}\\ \mathring{r}\mathring{\phi}
\end{pmatrix}
, \text{where } P\equiv
\begin{pmatrix}
    P_{tt} & P_{t\phi}\\
    P_{\phi t} & P_{\phi\phi}
\end{pmatrix}\label{eq:eqFromConserved}
\end{eqnarray}

When
\begin{eqnarray}
P^{-1}P'=
\begin{pmatrix}
\Gamma_{tr}^t&\Gamma_{r\phi}^t\\ \Gamma_{tr}^{\phi}&\Gamma_{r\phi}^{\phi}
\end{pmatrix}
\end{eqnarray}
is satisfied, Eq.\ref{eq:eqFromConserved} matches the $t$ and $\phi$ components of the geodesic equations.
When the metric is diagonal, it will reduce to the result for diagonal metric $P'/P=\text{diag}\{\Gamma_{tr}^t,\Gamma_{r\phi}^{\phi}\}$ since both $P$ and $\Gamma$ are diagonal. The conserved quantities $P(\mathring{t},\mathring{\phi})^T=(Q_t,Q_{\phi})^T$ give us $\mathring{t}$ and $\mathring{\phi}$ as functions of $r$: $(\mathring{t},\mathring{\phi})^T=P^{-1}(Q_t,Q_{\phi})^T\equiv (f(r),g(r))^T$. This allows us to convert the geodesic equation in terms of affine parameter $\lambda$ to time $t$ by using method shown in Sec.\ref{method:conversion}. Plug in the conversions $\mathring{r}=f\dot{r},\ringring{r}=f(f'\dot(r)^2+f\ddot{r}),\mathring{\phi}=f\dot{\phi},\ringring{\phi}=f(f'\dot{r}\dot{\phi}+f\ddot{\phi})$ into Eq.\ref{eq:nonDiag_r},\ref{eq:nonDiag_phi}, we arrive at
\begin{eqnarray}
\ddot{\phi}+\frac{2\dot{r}\dot{\phi}}{r}&=&c_{r\phi}\dot{r}\dot{\phi}-\Gamma_{tr}^{\phi}\dot{r}\\
\ddot{r}-r\dot{\phi}^2&=&c_{\phi\phi}\dot{\phi}^2-\Gamma_{t\phi}^r\dot{\phi}+c_0
\end{eqnarray}
where $c$'s are all functions of $r$ that $c_{\phi\phi}=(f'/f+\Gamma_{rr}^r)r^2-\Gamma_{\phi\phi}^r-r$, $c_0=-\Gamma_{tt}^r-(f'/f+\Gamma_{rr}^r)v^2$, $c_{r\phi}=-\Gamma_{r\phi}^{\phi}-f'/f+2/r$.

\subsection*{\textbf{E. Orbits around the horizon}}\label{method:aroundHorizon}
\begin{figure}[ht]
  \centering
  \includegraphics[width=0.8\textwidth]{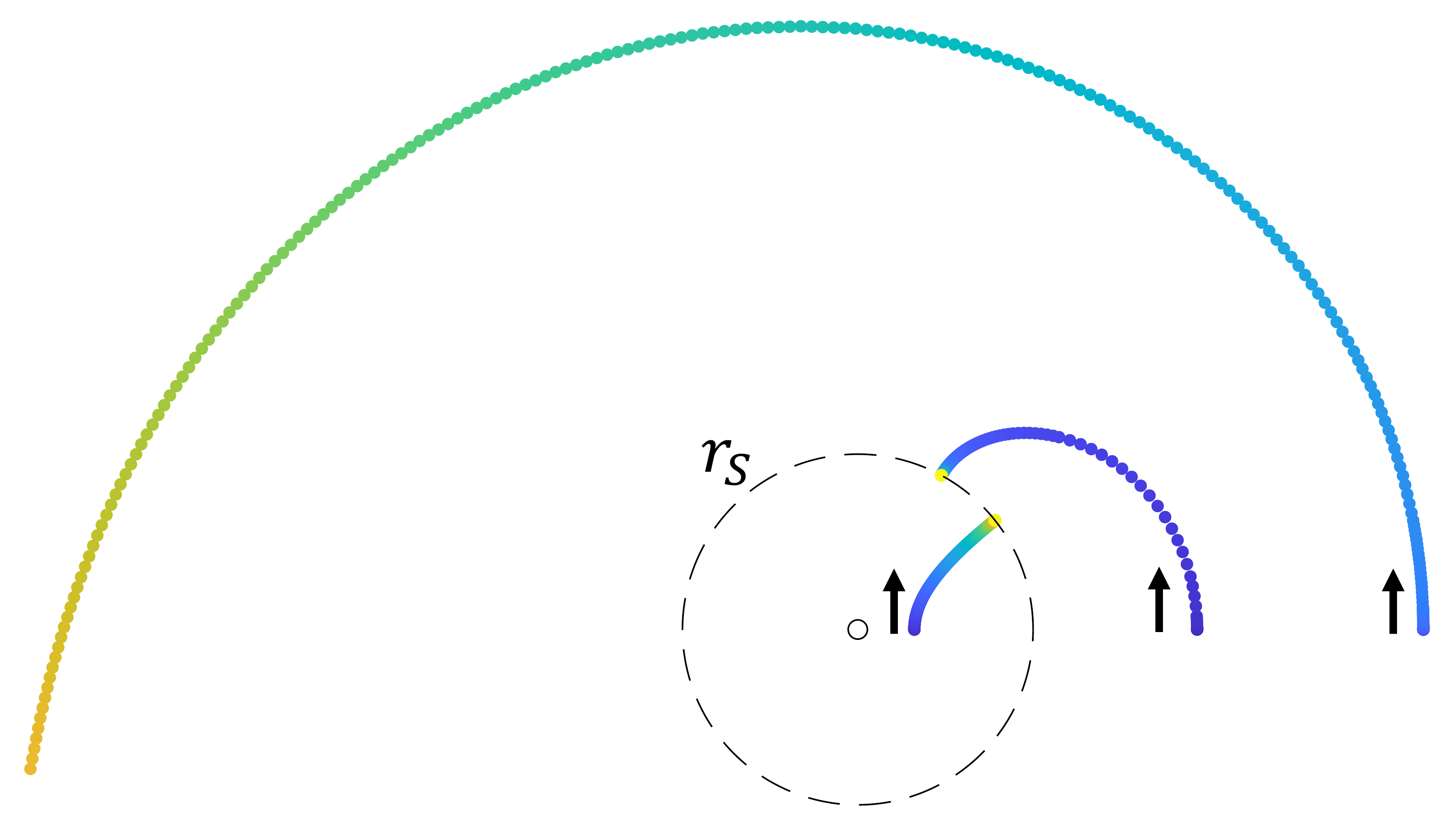}
  \caption{\textbf{Orbits around the horizon ($r_s=3.1$ mm).} As the initial radius gets closer to the event horizon shown in dashed line, the orbit approaches the blackhole and eventually is captured by the singularity at $r=r_s$. Using the method described earlier to integrate orbits outside the horizon, the orbit inside the horizon starts to show intriguing repulsion as predicted by \cite{maluf2016teleparallel}.}
  %\vspace{-5mm}
  \label{fig:aroundHorizon}
\end{figure}

We use the same model in previous section for Schwarzschild spacetime to integrate vehicle trajectory with controlled speed $v(r)$ and acceleration response to designed membrane elasticity $k(r)$. The integration uses
\begin{eqnarray}
r\ddot{\phi}+2\dot{r}\dot{\phi}&=&\left(\frac{v'}{v}+\frac{k}{v^2}\right)r\dot{r}\dot{\phi}\\
\ddot{r}-r\dot{\phi}^2&=&\left(\frac{v'}{v}+\frac{k}{v^2}\right)\dot{r}^2-k
\end{eqnarray}

which is derived from
\begin{eqnarray}
a_r=a_t\cos{\theta}-a_n\sin{\theta}\\
a_{\phi}=a_t\sin{\theta}+a_n\cos{\theta}
\end{eqnarray}
where the normal and tangential components of the acceleration are $a_n=k(r)\sin{\theta},a_t=dv/dt=(\partial v/\partial r) \dot{r}\equiv v' \dot{r}$. The orbits around and inside the event horizon are shown in Fig.\ref{fig:aroundHorizon}.

%\end{appendices}

%\bibliography{sn-bibliography}%

\newpage

\section*{References}
\begin{thebibliography}{<99>}
\bibitem{middleton2016elliptical} Middleton, C.A., Weller, D.: Elliptical-like orbits on a warped spandex fabric: A theoretical/experimental undergraduate research project. American Journal of Physics \textbf{84}(4), 284–292 (2016)

\bibitem{white2013trajectories} White, G.D.: On trajectories of rolling marbles in cones and other funnels. American Journal of Physics \textbf{81}(12), 890–898 (2013)

\bibitem{white2002shape} White, G.D., Walker, M.: The shape of “the spandex” and orbits upon its surface. American Journal of Physics \textbf{70}(1), 48–52 (2002)

\bibitem{aguilar2016review} Aguilar, J., Zhang, T., Qian, F., Kingsbury, M., McInroe, B., Mazouchova, N., Li, C., Maladen, R., Gong, C., Travers, M., et al.: A review on locomotion robophysics: the study of movement at the intersection of robotics, soft matter and dynamical systems. Reports on Progress in Physics \textbf{79}(11), 110001 (2016)

\bibitem{aydin2019physics} Aydin, Y.O., Rieser, J.M., Hubicki, C.M., Savoie, W., Goldman, D.I.: Physics approaches to natural locomotion: Every robot is an experiment. In: Robotic Systems and Autonomous Platforms, pp. 109–127. Elsevier, (2019)

\bibitem{fieldMediated} Li, S., Ozkan Aydin, Y., Xiao, C., Small, G., N. Gynai, H., Li, G., M. Rieser, J., Laguna, P., I. Goldman, D.: Field-mediated locomotor dynamics on highly deformable surfaces. Proceedings of the National Academy of Sciences \textbf{119}(30), e2113912119 (2022)
%arXiv preprint arXiv:2004.03549 (2021)

\bibitem{wheeler1991geons} Wheeler, J.A., Ford, K.W.: Geons, Black Holes, and Quantum Foam. Norton \& Company (2000)

\bibitem{middleton2014circular} Middleton, C.A., Langston, M.: Circular orbits on a warped spandex fabric. American Journal of Physics \textbf{82}(4), 287–294 (2014)

\bibitem{goldstein2002} Goldstein, Herbert and Poole, Charles and Safko, John: Classical mechanics. p24. Addison-Wesley (2002)

\bibitem{price2016spatial} Price, R.H.: Spatial curvature, spacetime curvature, and gravity. American Journal of Physics \textbf{84}(8), 588–592 (2016)

\bibitem{batlle2019exploring} Batlle, P., Teixid{\'o}, A., Llobera, J., Medrano, I., Pardo, L.C.: Exploring the rubber sheet spacetime analogy by studying ball movement in a bent trampoline. European Journal of Physics \textbf{40}(4), 045005 (2019)

\bibitem{drill} Gravitational Waves Work Like This Drill on Spandex, {https://www.youtube.com/watch?v=dw7U3BYMs4U}, Accessed: 2023-08-18

\bibitem{holmes2015teaching} Holmes, N.G., Wieman, C.E., Bonn, D.: Teaching critical thinking. Proceedings of the National Academy of Sciences \textbf{112}(36), 11199–11204 (2015)

\bibitem{smith2020direct} Smith, E.M., Stein, M.M., Walsh, C., Holmes, N.: Direct measurement of the impact of teaching experimentation in physics labs. Physical Review X \textbf{10}(1), 011029 (2020)

\bibitem{AydinBookChap} Aydin, Y.O., Rieser, J.M., Hubicki, C.M., Savoie, W., Goldman, D.I.: 6 - physics approaches to natural locomotion: Every robot is an experiment. In: Walsh, S.M., Strano, M.S. (eds.) Robotic Systems and Autonomous Platforms. Woodhead Publishing in Materials, pp. 109–127. Woodhead Publishing (2019)

\bibitem{unruh1981experimental} Unruh, W.G.: Experimental black-hole evaporation? Physical Review Letters \textbf{46}(21), 1351 (1981)

\bibitem{patrick2018black} Patrick, S., Coutant, A., Richartz, M., Weinfurtner, S.: Black hole quasibound states from a draining bathtub vortex flow. Physical review letters \textbf{121}(6), 061101 (2018)

\bibitem{steinhauer2016observation} Steinhauer, J.: Observation of quantum hawking radiation and its entanglement in an analogue black hole. Nature Physics \textbf{12}(10), 959 (2016)

\bibitem{zhu2018elastic} Zhu, J., Liu, Y., Liang, Z., Chen, T., Li, J.: Elastic waves in curved space: Mimicking a wormhole. Physical review letters \textbf{121}(23), 234301 (2018)

\bibitem{philbin2008fiber} Philbin, T.G., Kuklewicz, C., Robertson, S., Hill, S., K¨onig, F., Leonhardt, U.: Fiber-optical analog of the event horizon. Science \textbf{319}(5868), 1367–1370 (2008)

\bibitem{belgiorno2010hawking} Belgiorno, F., Cacciatori, S.L., Clerici, M., Gorini, V., Ortenzi, G., Rizzi, L., Rubino, E., Sala, V.G., Faccio, D.: Hawking radiation from ultrashort laser pulse filaments. Physical review letters \textbf{105}(20), 203901 (2010)

\bibitem{Barcelo2011} Barcel{\'o}, C., Liberati, S., Visser, M.: Analogue gravity. Living Reviews in Relativity \textbf{14}(1), 3 (2011)

\bibitem{couder2005dynamical} Couder, Y., Protiere, S., Fort, E., Boudaoud, A.: Dynamical phenomena: Walking and orbiting droplets. Nature \textbf{437}(7056), 208 (2005)

\bibitem{bush2015pilot} Bush, J.W.: Pilot-wave hydrodynamics. Annual Review of Fluid Mechanics \textbf{47}, 269–292 (2015)

\bibitem{BushQuantum} Bush, J.W.M.: Quantum mechanics writ large. Proceedings of the National Academy of Sciences \textbf{107}(41), 17455–17456 (2010)

\bibitem{tenev2018mechanics} Tenev, TG, Horstemeyer, MF: Mechanics of spacetime—A Solid Mechanics perspective on the theory of General Relativity. International Journal of Modern Physics D \textbf{27}(8), 1850083 (2018)

\bibitem{heng2014nature}  Heng, K.: The nature of scientific proof in the age of simulations. arXiv preprint arXiv:1404.6248 (2014)

\bibitem{ruphy2011limits} Ruphy, S.: Limits to modeling: Balancing ambition and outcome in astrophysics and cosmology. Simulation \& Gaming \textbf{42}(2), 177–194 (2011)

\bibitem{gruner2020computer} Gruner, S., Bartelmann, M.: On computer simulations, with particular regard to their application in contemporary astrophysics (2020)

\bibitem{maluf2016teleparallel} Maluf, Jos{\'e} Wadih: The teleparallel equivalent of general relativity and the gravitational centre of mass. Universe \textbf{2}(3), 19 (2016)

\bibitem{wang2021emergent} Wang, G., Phan, T.V., Li, S., Wombacher, M., Qu, J., Peng, Y., Chen, G., Goldman, D.I., Levin, S.A., Austin, R.H., et al.: Emergent field-driven robot swarm states. Physical review letters \textbf{126}(10), 108002 (2021)

%\bibitem{podolefsky2007refraining} Podolefsky, N.S., Finkelstein, N.D.: Refraining analogy: framing as a mechanism of analogy use. In: AIP Conference Proceedings, vol. 883, pp. 97–100 (2007). American Institute of Physics

%\bibitem{brown1987overcoming} Brown, D.E., Clement, J.: Overcoming misconceptions in mechanics: A comparison of two example-based teaching strategies. (1987)

%\bibitem{manogue2014tangible} Manogue, C.A., Gire, E., Roundy, D.: Tangible metaphors. In: Physics Education Research Conference Proceedings, pp. 27–30 (2014)


\end{thebibliography}

\end{document}